\documentstyle[preprint,tighten,aps,psfig]{revtex}
\begin{document}
\title{IS THERE DIQUARK CLUSTERING IN THE NUCLEON?}
\author{ L. Ya. Glozman$^1$ and K. Varga $^2$}
\address{ 
$^1$ High Energy Accelerator Research Organization (KEK),
Tanashi Branch, Tanashi, Tokyo 188-8501, Japan
\footnote{From September 1, 1999: Institute for Theoretical Physics,
University of Graz, Universit\"atsplatz 5, A-8010 Graz, Austria}\\
$^2$ Physics Division, Argonne National Laboratory, 
9700 South Cass Avenue, Argonne, Illinois 60638, USA
 \& Institute of Nuclear Physics of the Hungarian Academy of Sciences
(ATOMKI), Debrecen, PO BOX 51, Hungary}
\maketitle

\begin{abstract}
It is shown that the instanton-induced interaction in qq pairs,
iterated in t-channel, leads to a meson-exchange interactions
between quarks. In this way one can achieve a simultaneous understanding
of low-lying mesons, baryons and the nuclear force. The discussion 
is general and does not necessarily rely on the instanton-induced 
interaction. Any nonperturbative gluonic interaction between quarks, which
is a source of the dynamical chiral symmetry breaking and
explains the $\pi$ - $\rho$ mass splitting, will
imply an effective meson exchange picture in baryons. 
Due to the (anti)screening  there is a big difference 
between the initial 't Hooft interaction and the effective 
meson-exchange interaction.  It is demonstrated that
the effective meson-exchange interaction, adjusted to the 
baryon spectrum, does not bind the scalar diquark and does not
induce any significant quark-diquark clustering in the nucleon 
because of the nontrivial role played by the Pauli principle. 
\end{abstract}

\bigskip
\bigskip

PACS number(s): 12.38.Lg, 12.39.-x, 14.20.Dh

\section{Introduction}

Speculations that instantons could induce  diquark condensation 
in low temperature but high density quark matter \cite{RAPP,ALFORD}
have revived the interest in the diquark clustering in the nucleon.
It is  sometimes also argued that diquark condensation may occur even 
at  moderate densities, for example in heavy nuclei. This problem is 
strongly related to the question  of instanton induced  diquark clustering
in the nucleon. Indeed, the instanton-induced 't Hooft interaction
is strongly attractive for a quark-quark pair with quantum numbers
$T,J^P=0,0^+$ (scalar diquark). This raises expectations that
it binds a scalar diquark and
is responsible for the scalar diquark-quark structure of the nucleon
\cite{RAPP,SHURYAK}. This assumption is based on the iteration of
the 't Hooft interaction in the $qq$  s-channel. However, this
picture of the quark-quark interaction in baryons is only a small
part of a more general one, based on the effective meson-exchange
interaction \cite{GR,GLOZ}. When the 't Hooft interaction is first
iterated in the $qq$ t-channel it inevitably leads to  Goldstone
boson exchange between constituent quarks, which is drastically
different  from the initial (not iterated) 't Hooft interaction
due to  the (anti)screening effects.

The latter effective meson-exchange interaction does not induce
a bound scalar diquark, nor an appreciable diquark-quark
clustering in nucleon.
This effective meson exchange interaction
is also the most attractive  in $0,0^+$ $qq$ pairs, but the nature of this attraction
is very different  from that of the 't Hooft interaction.
This interaction, however, is not strong enough to bind the scalar
diquark. When it is combined with a confining interaction it binds
the diquark in the sense that there is no asymptotic state with
two free constituent quarks, though the mass of the scalar diquark
is a few tens of MeV above the two-constituent-quark threshold.
There is no significant diquark clustering in the nucleon either,
because the nucleon is intrinsically a three-quark system and the
fermionic-nature of the constituent quarks plays  an important  role.
If the subsystem of quarks 1 and 2 is in the $0,0^+$ state then
due to the antisymmetrization the quark pairs in the subsystems
1-3 and 2-3 are also partly in the $0,0^+$ state.
This implies that a strong attraction in $0,0^+$
quark pair contributes  in all quark subsystems simultaneously
and makes the nucleon compact, but without appreciable 
quark-diquark clustering.

This paper consists of two independent, but 
interrelated parts. In the first one we discuss 
how the instanton-induced interaction 
(or some general nonperturbative gluonic interaction)
leads to the poles when it is iterated in the
$qq$ t-channel. These pole contributions have an evident
meson-exchange interpretation. The latter meson-exchange
interaction is drastically different from the initial (bare)
't Hooft interaction which becomes strongly enhanced in the
channel of Goldstone boson exchange quantum numbers.

We also discuss the role of instantons in ${\bar q}q$
systems.
There is no new wisdom in that the nonperturbative
gluonic configurations, e.g. instantons, induce the dynamical
breaking of chiral symmetry  and explain the low-lying mesons.
We include the latter discussion only with the purpose of showing
how the nonperturbative gluonic interaction  both explains mesons
and  at the
same time  leads to the effective meson exchange
picture in the $qq$ systems. Through the latter it also explains  
the baryon spectra and the nuclear force. Our discussion is rather 
general, and does not necessarily rely on the instanton-induced 
interaction picture. Any nonperturbative gluonic interaction, 
which respects chiral symmetry and induces  the rearrangement
of the vacuum (i.e. dynamical breaking of chiral symmetry), will
automatically explain the $\pi - \rho$ mass splitting and will
imply a meson-exchange picture in baryons.

The second part of this paper is devoted to a detailed study
of diquark clustering in the nucleon, based on the effective
meson-exchange interactions in the baryons and the nucleon 
wave functions obtained from the solution of
the semirelativistic three-body Schr\"odinger equation. We show 
that there is no appreciable diquark clustering in the nucleon
and that the effective meson-exchange interaction, which is
adjusted to describe the baryon spectrum \cite{GPVW},  does not bind
the scalar diquark nor the nucleon. However, when this interaction
is combined with the confining interaction, one finds
a bound diquark
but with a mass above the two-quark threshold and 
very similar in magnitude to that 
obtained recently in lattice QCD  \cite{HES}.
Nevertheless, as soon as the strength of the effective meson-exchange
interaction is increased, not by a very big amount, it alone binds 
a nucleon, even without a confining force.
While the contributions from the confining interaction to the nucleon
mass are not small, 
the nucleon size, calculated with the confining interaction alone
and in a full model that includes both confinement and
effective meson exchange, is  different. It is substantially
smaller in the latter case, showing that there is indeed a soft interval
between the  scale when confinement becomes active, and the scale 
where  chiral physics starts to work. However, for excited baryon 
states, which are much bigger in  size, the role of confinement
increases.

\section{ How iteration in the t-channel
 of the instanton-induced interaction
leads to a meson-exchange picture and
 (anti)screens the short-range
behaviour.}

It has been shown in  recent years that a successful
explanation of light and strange baryon spectroscopy,
especially the correct ordering of the lowest states with
positive and negative parity, is
achieved if the hyperfine interaction between constituent
quarks $i$ and $j$ has a short-range behaviour
which reads schematically  \cite{GR}:

\begin{equation}
-\vec{ \lambda}_{i}^{F} 
\cdot \vec{\lambda}_{j}^{F}
\vec{\sigma}_i \cdot \vec{\sigma}_j,
\label{GBE}
\end{equation}

\noindent
where $\lambda^F$ is a set of a flavor Gell-Mann matrices
for $F=1,...,8$ and $\lambda^0 = \sqrt{2/3}{\bf 1}$. This
interaction is supplied by the short-range parts of 
Goldstone boson exchange (GBE) 
\footnote{ $\pi$, $K$
and $\eta$ exchanges; due to the axial anomaly the $\eta'$
is not a Goldstone boson, but in the large $N_c$ limit it
also becomes a Goldstone boson, and thus the coupling 
of $\eta'$ to a constituent quark should be essentially different
from that of octet mesons.},   
vector-meson-like exchange and/or correlated two-pseudoscalar-meson-like
exchange \cite{GLOZ}, etc.

It is  sometimes stated that the instanton-induced
't Hooft interaction in $qq$ pairs could also provide a good baryon
spectrum  as it contains 
a flavor- and spin-dependence and, iterated  in the $qq$
s-channel, produces a deeply bound scalar diquark which makes
the nucleon lighter than the $\Delta$ \cite{RAPP,SHURYAK}.
A similar picture of a deeply bound scalar diquark has been advocated
 in a generalized Nambu and Jona-Lasinio (NJL) model
\cite{VW,AR}. Then a baryon is constructed as an additive
diquark-quark system  or by solving ``relativistic
diquark-quark Faddeev equations'' that take into account the
quark exchange between the diquark and quark-spectator \cite{AR,BENTZ}. 
In this section we show that such a  picture
of baryons, based on the iteration of the local 4-fermion
interaction in the $qq$ s-channel is only a small part of a more
general picture, based on the meson-exchange interaction. The reason
is that when the 't Hooft interaction (or generalized NJL one)
 is first iterated in the $qq$ t-channel, it  inevitably leads to
the effective meson-exchange between constituent quarks, 
which is drastically
different  from the initial (not iterated) 4-fermion
local interaction due to (anti)screening effects. The difference is 
not only in the flavor- and spin-dependence, but  sometimes also in the 
sign of the interaction.

To demonstrate  this we  use a simple $2\times2$
't Hooft-determinant interaction
for two light flavors (u and d), neglecting 
for the simplicity the tensor coupling term, which is 
suppressed by the factor 
$\frac{1}{4N_c} = \frac{1}{12}$
 \cite{DIAKONOV}. We also assume  zero masses
for the current u and d quarks in this section.
 For our illustrative purposes 
such an approximation is justified. This Hamiltonian reads:

\begin{equation}
H=-G[(\bar{\psi}\psi)^2 +(\bar{\psi}i \gamma_5 \vec{\tau} \psi)^2
-(\bar{\psi} \vec{\tau} \psi)^2 - (\bar{\psi}i \gamma_5  \psi)^2].
\label{HOOFT}
\end{equation}

\noindent
The {\it dimensional} strength of the interaction $G$ as well as 
the ultraviolet 
cut-off scale $1/r_0$ can be related to parameters of the
instanton liquid \cite{DIAKONOV,SHURYAK}
(the dimensionless coupling constant is given by $G/4\pi r_0^2$). 
The interaction (\ref{HOOFT})
is attractive in the scalar-isoscalar $\bar{q} q$ channel (the first
term), leading to  chiral symmetry breaking, or, which is
related, to a  massive $\sigma$-meson field and
the constituent mass $m$ of quarks.  
This is readily obtained from the Schwinger-Dyson (gap)
equation for a quark Green function in the Hartree-Fock approximation.
The interaction in the $\bar{q}q$ pseudoscalar-isovector
channel is driven by the second term of (\ref{HOOFT}).
 It is so strong, that when it is iterated in the
$\bar{q}q$ s-channel by solving the Bethe-Salpeter equation,
see Fig. 1, it exactly compensates the $2m$-energy, supplied by the
first term in (\ref{HOOFT}), and thus there appear $T,J^P
=1,0^-$ mesons with zero mass as  deeply bound relativistic
$\bar{q}q$ systems - Nambu-Goldstone bosons. The nonzero mass
of the pseudoscalar mesons is brought about by the nonzero
 current quark mass as a perturbation, which is well
illustrated by the current algebra results
(Gell-Mann-Oakes-Renner relations). The first
two terms in the Hamiltonian (\ref{HOOFT}) form in fact
the classical NJL Hamiltonian \cite{NJL} and the statement
above is a theorem, proved by Nambu and
Jona-Lasinio many years ago. This scenario holds if the
fixed strength of the interaction $G$ exceeds some critical
level. In a more sophisticated derivation \cite{DIAKONOV}
the strength of the interaction $G$ is not fixed and should
be determined after one gets the chirally broken phase.

The Hamiltonian (\ref{HOOFT}) does not contain any
interaction in $\bar{q}q$ pairs with vector meson
quantum numbers. So, according to
the scenario above, the masses of vector mesons, $\rho$
and $\omega$, should be approximately $2m$, which is
well satisfied empirically. Thus, it cannot be overemphasized
that the $\pi-\rho$ mass splitting is brought about not by the
perturbative color-magnetic interaction between nonrelativistic
constituent quarks\footnote{It is also important  to remember
that the pion is not a simple  nonrelativistic two-body system,
but a purely relativistic 
${\bar q}q$ system and its
Nambu-Goldstone boson nature (zero mass in the chiral limit) 
cannot be obtained from a
nonrelativistic reduction of the second term in (\ref{HOOFT})
used in the Schr\"odinger equation.},
but by the detailed balance between the first and 
second terms in (\ref{HOOFT}), which is determined exclusively
by the demand that the   gluonic interaction between
current quarks must satisfy chiral $SU(2)_L \times SU(2)_R$
symmetry\footnote{The interaction (\ref{HOOFT}) is color-independent.
One can, of course, rewrite  this interaction
using the Pauli principle in terms of  linear combinations
of different operators, like ${\bf 1}$, ${\bf color \cdot color}$,
${\bf spin \cdot spin}$, ${\bf color \cdot color~ spin \cdot spin}$,
${\bf isospin \cdot isospin}$,... The presence of the 
${\bf color \cdot color~ spin \cdot spin}$ structure in this decomposition 
does not mean that
the interaction (\ref{HOOFT}) becomes similar in its effect 
to the color-magnetic
component  of the one gluon exchange interaction, which is 
explicitly color-dependent. It is like
an identity $a = (a-b) +b$ does not mean that effect of $b$
is contained in $a$.}.
 An important question, which is actively debated
nowdays, is which particular {\it nonperturbative} gluonic
configurations in QCD, e.g. instantons, or abelian monopoles,
or other topological configurations, are intrinsically
responsible for the chiral symmetry breaking.

Among other attractive features of the instanton-induced
interaction (\ref{HOOFT}) is that it automatically solves
the $U(1)_A$ problem, giving a much bigger mass to the
pseudoscalar flavor-singlet (in the present 2-flavor
formulation that is isosinglet) meson $\eta'$ \cite{THOOFT}.
This is because of the last term in (\ref{HOOFT}). Only
this term contributes  in a 
pseudoscalar flavor-singlet
quark-antiquark pair. Since this interaction is repulsive,
the $\eta'$ becomes heavy, contrary to $\pi$. We note in
passing that the color-magnetic interaction cannot explain
this big $\eta'-\pi$ mass splitting.

Clearly,  the simple Hamiltonian (\ref{HOOFT}) is only
some part of a more complicated physical situation. For
instance, one definitely needs some additionl attractive
interaction, e.g. confinement, otherwise the $\eta'$ meson
or vector mesons will be unbound, while the octet pseudoscalar
mesons are probably not affected  by the long-range
confining interaction.

Having mentioned all the positive features of the Hamiltonian
(\ref{HOOFT}) in the quark-antiquark system, we are now
going to discuss its implications in quark-quark systems,
i.e. in baryons. What is typically done is a Fierz-rearrangement
of the Hamiltonian (\ref{HOOFT}) into diquark $qq$ channels
\cite{RAPP,SHURYAK} (or, similarly, a Fierz-rearrangement
of the generalized NJL Hamiltonian into diquark channels
\cite{VW,AR}). Then the diquark Hamiltonian is iterated in the $qq$
s-channel, see Fig. 2. The interaction in the scalar
$T,J^P = 0,0^+$ diquark turns out to be attractive and
it produces a deeply bound scalar diquark\footnote{
In the first calculation
\cite{DFL} the scalar diquark was not bound for $N_c=3$.}.

However, as soon as the Hamiltonian (\ref{HOOFT}) is
iterated first in the $qq$ t-channel, see Fig. 3, it
implies  irreducible (for the $qq$ s-channel) pion- and
sigma-exchange interactions between quarks\footnote{We do not
show in Fig. 3 a lot of possible chains of bubbles which
would correspond to the irreducible two-meson-exchange with
crossed meson lines, three-meson-exchange, etc.}.
 This statement
comes about as a theorem since the iteration of the
Hamiltonian (\ref{HOOFT}) in $qq$ t-channel is equivalent
to its iteration in $\bar{q}q$ s-channel. Clearly 
the set of  diagrams in Fig. 3 contains all the diagrams
of Fig. 2, but in addition it contains  many others, 
and the effect of these additional diagrams
is so important that the physics implied by  Fig. 2
and Fig. 3 is drastically different\footnote{Sometimes
the Hamiltonian (\ref{HOOFT}) is applied in baryons
in the framework of the chiral quark-soliton model \cite{DPP}.
In this case  quark-quark correlations through
the self-consistent chiral {\it   mean field} and quantization of its
rotation take into account
 some part of the iterations in s-channel of
Fig. 2 and do not take into account the t-channel ladders of Fig. 3.}.

A simple example of the different physical implications
is that  Fig. 3 suggests a long-range meson-exchange
Yukawa tail, which is crucial for the interaction of
quarks, belonging to different nucleons, while if
the picture of Fig. 2 were correct the nuclear force
would be absent. Another evident difference is that
according to the Hamiltonian (\ref{HOOFT}) and Fig. 2
the interaction is absent in  flavor-symmetric, $T=1$,
quark pairs \footnote{I.e. the hyperfine 
interaction is absent in the $\Delta$-resonance and its excitations. 
If that were the case, the positive parity state $\Delta(1600)$,
which belongs to the $2\hbar\omega$ shell because of its
positive parity, would be approximately $\hbar\omega \simeq
500$ MeV above the negative parity pair $\Delta(1620) -
\Delta(1700)$.}, while the $qq$ interaction of Fig. 3
does not vanish in this case.

Less evident is that even the  ''short-range'' interaction
between quarks is crucially modified in Fig. 3
as compared to Fig. 2. We call it the ``(anti)screening effect''
and   illustrate it below \footnote{The subsequent qualitative
discussion will be extended and published in detail elsewhere 
\cite{PREP}.}.

In order to see it one should avoid the Fierz-rearrangement
of (\ref{HOOFT}) into a diquark Hamiltonian. Instead, one
can use the initial Hamiltonian (\ref{HOOFT}), but assume
that all initial, intermediate and final state $q_i q_j$
wave functions are explicitly antisymmetric. 

Consider
the first term of (\ref{HOOFT}). In the Nambu-Goldstone
mode of chiral symmetry a fermion field has a large
dynamical (constituent) mass $m$. Using a $1/m$
expansion, one obtains that to leading order ($m^0$)  
 the first term of (\ref{HOOFT}) leads to
a $\delta$-function type attraction in all quark pairs allowed  by Pauli
principle:

\begin{equation}
-G(\bar{\Psi}\Psi)^2 \Longrightarrow V({\vec r}_{ij})
=- 2G \delta({\vec r}_{ij}).
\label{FIRST}
\end{equation}

\noindent
The effect of the third term in (\ref{HOOFT}) to the same
order is

\begin{equation}
G(\bar{\Psi}\vec{\tau}\Psi)^2 \Longrightarrow V({\vec r}_{ij})
=2G \vec{\tau}_i \cdot \vec{\tau}_j \delta({\vec r}_{ij}).
\label{THIRD}
\end{equation}

\noindent
The potentials (\ref{FIRST}) and (\ref{THIRD}), combined
together, produce

\begin{equation}
 V({\vec r}_{ij})
=-2G (1-\vec{\tau}_i \cdot \vec{\tau}_j) \delta({\vec r}_{ij}).
\label{COMB}
\end{equation}

\noindent
Note that at this order the second and  fourth terms
of (\ref{HOOFT}) do not contribute. The potential (\ref{COMB})
suggests a strong attraction in the isospin-zero quark pair,
and no interaction  in $T=1$ quark pairs. Assuming relative
angular momentum $L=0$ within the $T=0$
quark pair the Pauli principle implies that the spins of the quarks
should be antiparallel, $S=0$. When the theory is sensibly
regularized the delta-function attraction is smeared
out over the instanton size $r_0$

\begin{equation}
\delta(\vec r)\rightarrow \frac{1}{4\pi r_0^2}
\frac{e^{- r/r_0}}{r}.
\label{DELTA}
\end{equation} 

\noindent
This substitution arises from a replacement of the static
Green function of the infinitely heavy particle in (\ref{HOOFT})

\begin{equation}
G_{\mu = \infty}({\vec x} - {\vec y}) = - r_0^2 \delta
({\vec x} - {\vec y}),
\label{heavy}
\end{equation}

\noindent
by the Green function of a particle with mass $\mu = 1/r_0$

\begin{equation}
G_{\mu = 1/r_0}({\vec x} - {\vec y}) = - \frac{1}{4\pi}
\frac{e^{-\mid{\vec x} - {\vec y}\mid/r_0}}{\mid{\vec x} - {\vec y}\mid}.
\label{INST}
\end{equation}

When the strength of the interaction is big enough, the
potential (\ref{COMB})-(\ref{DELTA}), iterated by solving the 
semirelativistic Schr\"odinger
equation (i.e. when the kinetic energy operator is taken in
a relativistic form) can produce a deeply bound scalar diquark, in agreement
with \cite{RAPP,SHURYAK}. Indeed, when one takes the strength
$G=490\frac{1}{8N_c^2}$ GeV$^{-2}$, $N_c =3$, 
with the instanton size
$r_0$ between 0.3 and 0.35 fm  and the constituent
mass $m=340-400$ MeV \cite{RAPP,SHURYAK} one  finds a very deeply bound diqurk.

 In the illustration above we have used
a simplified but transparent nonrelativistic picture that
adequately reflects in the present case the essential features 
of a more rigorous Bethe-Salpeter approach.

Sometimes the potential (\ref{COMB}) is applied to explain
the hyperfine splittings in baryons \cite{SR,METSCH}. While
it can  generate the $\Delta - N$ mass splitting,
it fails to explain  the lowest levels with positive and
negative parity because it does not contain the necessary
spin-isospin dependence (\ref{GBE}). It will become evident from
the discussion below that such an interpretation of the role of
instantons in baryons does not survive as soon as the wider
class of diagrams in Fig. 3 is considered.

What happens  when the first term in (\ref{HOOFT}) is
iterated in the $qq$ t-channel? The corresponding 
amplitude is

\begin{equation}
T_S(q^2) = 2G + 2G J_S(q^2) 2G +... = \frac{2G}{1-2G J_S(q^2)},
\label{ITER}
\end{equation}

\noindent
where $J_S(q^2)$ is the  loop integral
(bubble) with the scalar vertex which represents vacuum
polarization in the scalar channel. The eq. (\ref{ITER})
defines ``running amplitude'' and a negative sign in the
denominator implies its antiscreening behaviour.
The expression  
(\ref{ITER}) is known to have a pole at $q^2 = 4m^2$
\cite{NJL}, which can be identified with the exchange by
scalar meson $\sigma$ with the mass $\mu_\sigma = 2m$
in the chiral limit. The coupling constant of the $\sigma$-meson
to constituent quark can be obtained as a residue of
(\ref{ITER}) at the pole

\begin{equation}
\frac{g_{\sigma q}^2 }{q^2-\mu_\sigma^2} = -\frac{2G}{1-2G J_S(q^2)}.
\label{POLE}
\end{equation}

\noindent
 Expanding
 the $\bar \Psi \sigma \Psi$ vertex in $1/m$, one obtains
to leading order $(m^0)$ the following well-known sigma-exchange
potential

\begin{equation}
 V_\sigma({ r}_{ij})
=-\frac{g^2_{\sigma q}}{4\pi} \frac{e^{-\mu_\sigma { r}_{ij}}}
{{r}_{ij}}.
\label{SIGMA}
\end{equation}

 The equivalence
between the t-channel ladder of bubbles in Fig. 3
beyond the $\sigma$-meson pole in the t-channel and
the meson-exchange diagram is achieved  
only when some form factor $F_{\sigma q}(q^2)$ is inserted 
into the meson-quark vertex, i.e. the left hand side
of eq. (\ref{POLE}) should be multiplied with
$F^2_{\sigma q}(q^2)$. The form factor is to be normalized
$F_{\sigma q}(q^2= \mu_\sigma^2) = 1$.

 In principle the eq. (\ref{POLE}) allows to obtain
a functional form for such a form factor. However, it will
be very far from reality because the toy model (\ref{HOOFT})
does not contain confinement, which should be important
for the interaction between quark and antiquark in the
weakly bound system like $\sigma$-meson (note that its mass
is just at the ``continuum threshold'' $2m$). In this situation
the best way is to rely on our general understanding of
the low-energy effective theory. Both constituent quarks
and chiral meson fields as well as their couplings make a sense
only in the Nambu-Goldstone mode of chiral symmetry. 
When momentum transfer at the meson - quark vertex
exceeds the chiral symmetry breaking scale 
$\Lambda_\chi$ ( which within all NJL-like models
coincides with the regularization scale $1/r_0$)
the effective theory should be
cut off. This cut off is accomplished by a form factor
in the meson - constituent quark vertex and should
be related to the internal structure of both quasiparticles.
But in any case the scale parameter in this form factor should 
be comparable with $\Lambda_\chi$. At high momenta one can
use neither constituent quark nor chiral fields and original
quark-gluon degrees of freedom should be used instead.

Approximating this form factor by

\begin{equation}
F_{\sigma q}^2(q^2) = 
\frac{\Lambda^2_\sigma - \mu_\sigma^2}{\Lambda^2_\sigma - q^2},
\label{FORM}
\end{equation}
 instead of the potential (\ref{SIGMA}) we arrive at

\begin{equation}
 V_\sigma({ r}_{ij})
=-\frac{g^2_{\sigma q}}{4\pi}\left( \frac{e^{-\mu_\sigma { r}_{ij}}}
{{r}_{ij}} -  \frac{e^{-\Lambda_\sigma { r}_{ij}}}
{{r}_{ij}}\right).
\label{SIGFOR}
\end{equation}

\noindent
Note, that any  functional form of form factor 
leads to a similar suppression of the potential at
short range but it does not influence its long-range
part which is determined exclusively by the position
of the pole. If one takes a dipole form factor 
the suppression will be stronger.

What is the  fate of the third term in (\ref{HOOFT}), when it is
iterated in the $qq$ t-channel? In this case one obtains
the following amplitude

\begin{equation}
T_S^{ab}(q^2)= -\left( 2G - 2G J_S(q^2) 2G +...\right) \delta_{ab}
= - \frac{2G}{1+2G J_S(q^2)} \delta_{ab},
\label{ITER2}
\end{equation}

\noindent
where $a,b$ are isospin indices.
The positive sign in the denominator indicates screening. For
instance, at $q^2 = 4m^2$ the strength of the interaction is
reduced by the factor 2 versus a bare vertex. Still, this
suppression of the interaction at low momenta is not realistic,
because the toy model (\ref{HOOFT}) does not contain confinement
and thus there is only a repulsion in the scalar-isovector quark-antiquark
system.
When confinement is added in the quark-antiquark pairs, there appear
heavy scalar-isovector mesons and the sign of the
amplitude (\ref{ITER2}) becomes opposite at small momenta! This
low-momentum amplitude corresponds to the exchange by
scalar-isovector mesons between quarks.
The corresponding meson-exchange interaction is
 similar in form to (\ref{SIGFOR}), but with an additional
factor $\vec \tau_i \cdot \vec \tau_j$.   The expectation value of
the operator $\vec{\tau}_i \cdot \vec{\tau}_j$
in the scalar diquark is
$-3$  and thus the interaction (\ref{COMB})
is stronger by the factor 4 through the interaction (\ref{THIRD})
in the picture of Fig. 2. In contrast, in the picture of Fig. 3 the
contributions from the scalar and scalar-isovector meson exchanges
tend to cancel each other.

Thus we see that the initial interaction is screened.
This screening means that  the interaction
(\ref{COMB}) becomes  weaken and that
its isospin dependence is modified.
It is  trivial to check that the attraction (\ref{SIGFOR})
does not lead to a bound scalar diquark with any reasonable coupling
constant, sigma-meson mass, cut-off mass $\Lambda_\sigma$ and 
constituent quark mass (see discussion in the next chapter). 
The scalar-isovector meson exchange
will further reduce this attraction, though the coupling
constant of the scalar-isovector mesons to constituent quarks
will be essentially smaller.

Both the scalar-isoscalar exchange and scalar-isovector
exchanges between constituent quarks do not contain the
 flavor-spin dependence  (\ref{GBE})
 which is necessary for baryon spectroscopy\footnote{ 
We do not state, however, that such an interaction
is unimportant. Just the opposite, the $\sigma$-exchange is known
to be very important for the medium-range attraction in the NN
system and it also contributes to binding nucleon.
 However it only has  
a small influence on the splittings via different radial behaviour
of  baryon wave functions.}.

Now we shall extend our $1/m$ expansion of the Hamiltonian
(\ref{HOOFT}) to the next-to-leading order, taking into
account terms $m^{-2}$. The first and  third terms of
(\ref{HOOFT}) will give at this order the spin-orbit forces,
as well as some small corrections to the interactions
(\ref{FIRST}) and (\ref{THIRD}). 
 The second and  fourth terms generate, however,
a flavor-spin dependent interaction. Consider the second
term in (\ref{HOOFT}). It gives both the spin-spin and
tensor force components. We ignore below the tensor force
as it is irrelevant to our simple discussion and
for the $L=0$ $qq$ pair. Then:

\begin{equation}
-G(\bar{\Psi}i\gamma_5\vec{\tau}\Psi)^2
\Longrightarrow V({\vec r}_{ij})
=2G \frac{1}{12m^2}\vec{\sigma}_i \cdot \vec{\sigma}_j
\vec{\tau}_i \cdot \vec{\tau}_j \Delta\delta({\vec r}_{ij}).
\label{SECOND}
\end{equation}

\noindent
Again, assuming that the instanton has a finite size $r_0$,
the potential (\ref{SECOND}) reads:

\begin{equation}
 V({\vec r}_{ij})
=\frac{2G}{4\pi r_0^2} \frac{1}{12m^2}\vec{\sigma}_i \cdot \vec{\sigma}_j
\vec{\tau}_i \cdot \vec{\tau}_j
\left(\frac{1}{r_0^2} \frac{e^{-r_{ij}/r_0}}{r_{ij}} - 
4\pi\delta({\vec r}_{ij})\right).
\label{SECMOD}
\end{equation}

Let us now iterate the second term of (\ref{HOOFT}) in the $qq$
t-channel

\begin{equation}
T_P^{ab}(q^2) = \left(2G + 2G J_P(q^2) 2G +...\right)\delta_{ab} = 
\frac{2G}{1-2G J_P(q^2)} \delta_{ab},
\label{ITERP}
\end{equation}

\noindent
where $J_P(q^2)$ is a bubble  with a pseudoscalar vertex
(vacuum polarization in the pseudoscalar channel).
There is a pole at $q^2=0$
 in (\ref{ITERP}) in the chiral
limit, which can be identified as a pion-exchange
(beyond the chiral limit it is shifted
to a physical pion mass $q^2=\mu_\pi^2$.) The coupling constant
of pion to constituent quark can be obtained as a residue
of (\ref{ITERP}) at the pole

\begin{equation}
\frac{g_{\pi q}^2 }{q^2} = -\frac{2G}{1-2G J_P(q^2)}.
\label{POLEP}
\end{equation}

\noindent

Thus near the pole (\ref{ITERP}) - (\ref{POLEP}) represents
a pion-exchange potential between quarks, which
in the chiral limit, $\mu_\pi =0$, at the order $1/m^2$
(omitting the tensor force component) 
is:

\begin{equation}
V_\pi({\vec r}_{ij})
=-\frac{g_{\pi q}^2}{4\pi} \frac{1}{12m^2}\vec{\sigma}_i \cdot \vec{\sigma}_j
\vec{\tau}_i \cdot \vec{\tau}_j 4\pi\delta({\vec r}_{ij}).
\label{PION}
\end{equation}

The difference between (\ref{SECOND}-\ref{SECMOD}) and (\ref{PION}) is obvious:
the interaction (\ref{PION}) is much stronger.\footnote{One can easily see
it from comparison of $<\Psi_0 | \Delta \delta(\vec r) | \Psi_0>$
and $<\Psi_0 |  \delta(\vec r) | \Psi_0>$ where $\Psi_0$
is a zero order function stemming from confinement.} A source of
this enhancement is also obvious: near the pole the original bare 
interaction
$2G$ becomes strongly reinforced \footnote{In essense this 
antiscreening is some kind of asymptotic freedom: at space-like momenta
$q^2 \rightarrow -\infty$ the interaction is represented by
a bare vertex $2G$, but at $q^2 \rightarrow 0$ it becomes
infinitely enhanced in the channel with Goldstone boson exchange
quantum numbers.}. The pion pole is located just
near the space-like region and thus strongly influences the
quark-quark interaction at not very high momentum transfer.

 Again, to retain
the equivalence between the t-channel ladder of bubbles
in Fig. 3 and the pion-exchange diagram beyond the pole,
one must insert a form factor into the $\pi q$ vertex. The effect
of this form factor is to smear out the $\delta$-type
interaction in (\ref{PION}) over the region $1/\Lambda_{\pi q}$. 
If this form factor is choosen in the form (\ref{FORM}),
then one obtains

\begin{equation}
V_\pi({\vec r}_{ij})
=-\frac{g_{\pi q}^2}{4\pi} \frac{1}{12m^2}\vec{\sigma}_i \cdot \vec{\sigma}_j
\vec{\tau}_i \cdot \vec{\tau}_j \Lambda_{\pi q}^2
\frac{e^{-\Lambda_{\pi q}r_{ij}}}
{r_{ij}}.
\label{PIONSM}
\end{equation}

The $m^{-2}$  expansion of the fourth term of 
the Hamiltonian (\ref{HOOFT}) will give a result similar to
second term, without, however, isospin-dependent
factor and with the opposite sign.
Its iteration
in the $qq$ t-channel will produce screening
effects as it is repulsive in the ${\bar q} q$ s-channel. When,
however, this term is combined with an additional
attractive interaction, e.g. confinement, it will give
$\eta'$. Then the iteration in the $qq$ t-channel will
imply  $\eta'$-exchange between quarks. The latter
interaction is similar to (\ref{PION}), except that
the factor ${\vec \tau}_i \cdot {\vec \tau}_j$ is not
present and that in this case there appears a
Yukawa part of the potential because the mass of $\eta'$
is not zero in the chiral limit.

 The discussion above  suggests
that while for the picture of Fig. 2 the most important interaction
is (\ref{COMB}) and the interaction (\ref{SECOND}) is only some
very small correction to it, in the case of Fig. 3 the most
important interaction in baryons becomes (\ref{PION}) and the
one of (\ref{SIGFOR})  only plays a  modest role for splittings.
This is a consequence of an antiscreening. The antiscreening implies that
if a typical momentum transfer in the meson-quark vertex
(which in qq systems is of the same order as momentum of quarks)
is below the chiral symmetry breaking scale, then 
 the original (bare) quark-quark vertex in the pseudoscalar channel is
strongly reinforced by the pole 
which occurs when one iterates it in the t-channel.
These pole contributions represent the Goldstone boson exchange
interactions between constituent quarks in the Nambu-Goldstone
mode of chiral symmetry. Only at a rather high momentum
transfer (i.e. very far from the poles) there should appear 
a sensitivity to the original (bare)
quark-quark vertex. In the latter case the constituent
quarks and chiral fields cannot be used as effective
degrees of freedom. So the crucial question is what  a
typical momentum transfer in the given system is. In the low-lying
baryons it is below the chiral symmetry breaking scale
thus justifying a use of the effective qq interactions there
\cite{GL}.

We hope that the discussion above has been transparent 
enough to show a dramatic difference between the initial
't Hooft interaction, taken literally in $qq$ system,
and its implication after iteration in the t-channel, producing
meson exchange between constituent quarks. 

In fact, what one needs for the chiral symmetry breaking
is a scalar interaction between quarks. Any pairwise
gluonic interaction between quarks in the local approximation
will necessarily contain the first and second terms of (\ref{HOOFT})
with fixed relative strength. This is because of the chiral
invariance. Thus all our conclusions are rather general and
do not rely necessarily on 't Hooft interaction. 
An important
lesson is to see how this nonperturbative gluonic interaction,
which induces the dynamical breaking of chiral symmetry, suggests
an explanation of both the low-lying mesons and at the same time
of baryons and the nuclear force through the effective meson exchange
picure in $qq$ systems. 
Among the various applications of this idea, will be to work out
how the meson-exchange interaction  shifts the transition
point from the chiral symmetry broken phase to the 
color-superconductor phase.

We also mention a recent lattice study \cite{LIU} which shows
directly that the hyperfine $\Delta - N$ splitting is mostly
due to the meson-exchange interaction between quarks. Another
indirect evidence in favor of the picture in Fig. 3 versus that
 in Fig. 2 is that after cooling (the cooling means that
all gluonic configurations, except for instantons, are
removed) the $\Delta -N$ splitting disappears \cite{NEGELE}.
While the cooling does not affect the initial 't Hooft
interaction between quarks and thus the whole s-channel
ladder of Fig. 2 is active, it ruins the t-channel ladder
of Fig. 3. The reason is that there are not enough antiquarks
in the Fock space after cooling as  in quenched approximation 
they are mostly produced by different gluons, including
perturbative ones, attached to valence quark lines (Z graphs).

\section{Are the diquark and nucleon bound by the meson-exchange
interaction?}

We start this section with a short description of the effective
meson-exchange interaction model, adjusted
to describe baryon spectroscopy within an exact semirelativistic
3-body formulation \cite{GPVW}. The 
Hamiltonian of  ref. \cite{GPVW} reads:

\begin{equation}
\label{htot}
H=\sum_{i=1}^3\sqrt{\vec{p}_i{}^2+m_i^2}+\sum_{i<j=1}^3V_{ij},
\end{equation}

$$ \sum_{i=1}^3{\vec p}_i = 0.$$

\noindent
Here the relativistic form of the kinetic-energy operator is used, with
$\vec{p}_i$ the 3-momentum and $m_i$ the masses of the constituent quarks. The
dynamical part consists of the quark-quark interaction

\noindent
\begin{equation}
V_{ij}=V_{\chi}+V_{conf}.
\end{equation}

\noindent
The linear pairwise confining potential 

\begin{equation}
V_{conf}(r_{ij})=V_0+C r_{ij},
\label{conf}
\end{equation}

\noindent
\noindent
includes both the color-electric string $Cr_{ij}$ with the
color factor absorbed into the string tension $C$ as well as a constant
$V_0$, which is large and negative, and
thus effectively includes all possible spin- and flavor-independent
attractive interactions between quarks, e.g. $\sigma$-exchange
(\ref{SIGFOR}), etc. The flavor- and spin-dependent part of
the above Hamiltonian is

\begin{eqnarray}
\label{voct}
V_\chi(\vec r_{ij})  &=&
\left[\sum_{F=1}^3 V_{\pi}(\vec r_{ij}) \lambda_i^F \lambda_j^F\right.
+\sum_{F=4}^7 V_K(\vec r_{ij}) \lambda_i^F \lambda_j^F
\nonumber \\ 
&&\left.\raisebox{0ex}[3ex][3ex]{}+V_{\eta}(\vec r_{ij}) \lambda_i^8 
\lambda_j^8 + \frac{2}{3}V_{\eta'}(\vec r_{ij})\right]
\vec\sigma_i\cdot\vec\sigma_j,
\end{eqnarray}

\begin{equation}
\label{vyuk}
V_\gamma (\vec r_{ij})= \frac{g_\gamma^2}{4\pi}
\frac{1}{12m_im_j}
\left\{\mu_\gamma^2\frac{e^{-\mu_\gamma r_{ij}}}{ r_{ij}}-
\Lambda_\gamma^2\frac{e^{-\Lambda_\gamma r_{ij}}}{ r_{ij}}\right\}.
\end{equation}

\noindent
with $\mu_\gamma$ ($\gamma=\pi,K,\eta,\eta'$) being the individual 
phenomenological meson masses, and
$g_\gamma^2/4\pi$ the meson-quark coupling constants.

The constituent mass of the light quarks $m=m_u=m_d$ was fixed
in \cite{GPVW} to a typical value, $m=340$ MeV, implied
by a simple static quark model formula for the nucleon magnetic
moment.  It is astonishing  that the same value has
been obtained in a lattice measurement \cite{HES}. All
other parameters of the above Hamiltonian can be found in
ref. \cite{GPVW}.

In  light quark systems, like $N$ and $\Delta$, only the
$\pi$-like, $\eta$-like and $\eta'$-like parts of the
 potential (\ref{voct}) contribute. The $\pi$-like exchange
interaction is determined by the following matrix elements:

\begin{equation}
-\vec\tau_i\cdot\vec\tau_j \vec\sigma_i\cdot\vec\sigma_j
 = \left\{ \begin{array}{llll}
-1, & {\rm if} \; S_{ij} =1, & T_{ij} =1 \\
-9, & {\rm if} \; S_{ij} =0, & T_{ij} =0 \\
 3, & {\rm if}\; S_{ij} =1, & T_{ij} =0 \\
 3, & {\rm if}\; S_{ij} =0, & T_{ij} =1, 
\end{array} \!\!\right. 
\label{st}
\end{equation}

\noindent
while the $\eta$- and $\eta'$-like exchanges depend only
on the spin $S_{ij}$ of a quark pair.

\noindent
Combining all $\pi$, $\eta$ and $\eta'$ interactions
one finds that the potential (\ref{voct}) is most attractive at
short distances in $S_{ij},T_{ij}=0,0$ quark pair and essentially
less attractive in the $S_{ij},T_{ij}=1,1$ diquark system. In other
possible color-antitriplet $qq$ pairs it is repulsive.

Applying the Hamiltonian (\ref{htot}) in a color-antitriplet
$qq$ system, one finds a mass $m_{00}= 744$ MeV for a
scalar diquark, $T,J^P=0,0^+$, and a mass
$m_{11}= 869$ MeV for an
axial-vector diquark, $T,J^P=1,1^+$. In both cases
the relative orbital angular momentum is $L=0$, so
the total angular momentum coinsides with the spin
of two quarks. These values are very similar to
those obtained recently from the lattice
``diquark spectroscopy'' \cite{HES}.
The root-mean-square (r.m.s.) radius of the scalar diquark
is 0.354 fm and of the axial-vector one - 0.438 fm.
These radii do not include the size of the constituent
quark.

It is evident that the confining interaction 
implies a bound diquark in the sense that there are no 
asymptotically free constituent quarks. So it is very 
instructive to compare the mass of the above diquarks 
with the unphysical two-constituent-mass threshold, 
$2m=680$ MeV. The scalar
diquark mass is a few tens of MeV above the threshold,
which indicates that the meson-exchange part of the
interaction, including $V_0$, does not bind a diquark
without the confining interaction $Cr_{ij}$.
This can also be checked explicitly. To this end 
we combine the spin- and isospin-dependent interaction (\ref{voct})
with the $\sigma$-exchange potential (\ref{SIGFOR})
and drop the confining potential (\ref{conf}).
The $\sigma q$ coupling constant is constrained
to be equal to the $\pi q$ one, as suggested by chiral symmetry,

\begin{equation}
\frac{g^2_{\sigma q}}{4\pi} = \frac{g^2_{\pi q}}{4\pi} = 0.67.
\end{equation} 

\noindent
The sigma mass is taken to be $\mu_\sigma  \simeq  2m$,
which is implied by the well known result for all
NJL-like interactions,  $\mu^2_\sigma = 4m^2 + \mu_\pi^2$.
With these constraints we do not find a bound diquark
with any reasonable value for $\Lambda_\sigma \sim 1$ GeV.
Only with  $\Lambda_\sigma > 3$ GeV does a weakly
bound scalar diquark appear. If one increases the 
$\frac{g^2_{\sigma q}}{4\pi}$
coupling constant by a factor 1.5, but keeps the $\pi q$
coupling constant,
then we obtain a bound
diquark only at  $\Lambda_\sigma > 1.6$ GeV.
Thus we conclude that the meson-exchange interaction itself
does not bind a diquark.

The next question we address in this section is whether the
meson-exchange interaction binds nucleon itself, without
confinement. A-priori one cannot exclude the possibility
that while the diquark
is unbound the three quark system will be bound because of
genuine 3-body effects (compare, e.g.,
the binding energy of tritium and deuteron). Indeed,
a full model, including confinement (\ref{conf}), produces a
nucleon mass which is below the three constituent mass threshold,
$3m=1020$ MeV. Hence the positive contribution from the rising
potential $Cr_{ij}$, 631 MeV, is not big  compared
to the negative contribution
from $3V_0=-1248$ MeV in combination with the 
negative contribution of the
spin-dependent part of interaction, -750 MeV.

Superficially one could thus conclude that the meson-exchange part of
the Hamiltonian could bind nucleon without any support from confinement.
However, such an interpretation  cannot be taken
for two reasons. Firstly, we do not know which
part of the negative constant $V_0$ comes from the $\sigma$-exchange,
and which - from the genuine color-electric confinement, because the
$Y$-shape of the gauge-invariant 3-body confining interaction
can be approximated by a sum of  pairwise potentials only when
some additional constant contribution is added. Secondly, a perfect
fit of the baryon spectrum with a quality similar to that 
in ref. \cite{GPVW} can be obtained with a constituent mass
smaller than $m_N/3$. So we have performed a direct calculation
of the nucleon, replacing the potential (\ref{conf}) by the
$\sigma$-exchange potential  (\ref{SIGFOR}). We have found that
for  $\Lambda_\sigma \leq 1800$ MeV the nucleon is unbound and becomes
bound at higher values of $\Lambda_\sigma$. If one increases the
$\frac{g^2_{\sigma q}}{4\pi}$ by a factor 1.5, then the nucleon
becomes bound for   $\Lambda_\sigma > 1200$ MeV. These results
indicate that while the nucleon is unbound with the meson-exchange
potential parameters fixed of ref. \cite{GPVW}, it could be bound
as soon as a spin- and isospin-independent $\sigma$-like
exchange interaction and/or spin- and isospin-dependent interactions
are made stronger, not by a big amount. 
It is  evident that a description
of all excited states demands the presence of  confinement
because all these states are much above the $3m$ threshold.

\section{Is there diquark clustering in the nucleon?}

 We shall use the following
set of Jacobi coordinates and a coupling scheme with self-evident notation:

\begin{equation}
\vec \rho =  {\vec r}_1 -  {\vec r}_2; ~ 
\vec \lambda =  {\vec r}_3 - \frac{ {\vec r}_1 + {\vec r}_2}{2},
\end{equation}
 
\begin{equation}
{\vec S}_{12} = {\vec S}_{1} + {\vec S}_{2}; ~ 
{\vec S} = {\vec S}_{12} + {\vec S}_{3},
\end{equation}

\begin{equation}
{\vec L} = {\vec L}_\rho + {\vec L}_\lambda,  
\end{equation}

\begin{equation}
{\vec J} = {\vec L} + {\vec S},  
\end{equation}

\begin{equation}
{\vec T}_{12} = {\vec T}_{1} + {\vec T}_{2}; ~ 
{\vec T} = {\vec T}_{12} + {\vec T}_{3}.
\end{equation}

Let $P_{S_{12}T_{12}}$  be a projector onto a
subspace with a given value of spin $S_{12}$ and
isospin $T_{12}$ of the particles 1 and 2. The probability
density for finding particles 1 and 2 in a spin-isospin state
$S_{12}T_{12}$ at a relative distance $r_{12}$ is given
by

\begin{equation}
g_{S_{12}T_{12}}(r_{12}) = 
< \Psi |P_{S_{12}T_{12}} \delta(\rho - r_{12}) | \Psi>,
\end{equation}

\noindent
where $\Psi$ is an antisymmetric 3-body baryon wave function. 
One can similarly
define the probability density for finding  the particle 3
at a
distance $r_{12,3}$ 
from the center of mass of particles 1 and 2

\begin{equation}
f_{S_{12}T_{12}}(r_{12,3}) = 
< \Psi |P_{S_{12}T_{12}} \delta(\lambda - r_{12,3}) | \Psi>.
\end{equation}

\noindent
Then one can calculate the corresponding moments 

\begin{equation}
<r_{12}^k> = \int dr_{12} r_{12}^k g_{S_{12}T_{12}}(r_{12}),
\end{equation}

\begin{equation}
<r_{12,3}^k> = \int dr_{12,3} r_{12,3}^k f_{S_{12}T_{12}}(r_{12,3}).
\end{equation}

In  Table 1 we present the $k=2$ moments  
for $N$ and $\Delta$ in two cases: (i) full model,
(ii) no spin-dependent
interaction at all (i.e. only confinement is active).

Comparing the nucleon r.m.s. radius, $\sqrt{<r_N^2>}$ = 0.304 fm,
with the radius of a scalar diquark, 0.354 fm, we can deduce 
the role of genuine 3 body effects - they make the nucleon essentially
more compact than the diquark.

The empirical mean square charge radius of the proton, $0.86^2$ fm$^2$,
consists of a few contributions: the contribution from the mean square
matter radius above, the charge mean square radius of the constituent
quark, the meson exchange current contribution \cite{HELM,BAR}, the proton
anomalous magnetic moment contribution, etc. The rather small
value of the matter radius, obtained above, is consistent with large
contributions from other sources. For instance, the charge radius
of the constituent quark should be mainly determined by the $\rho$-meson
pole in the time-like region (vector meson dominance) and thus
can be expected to be of the order $\sim 0.6$ fm. 

The r.m.s. radius of the $\Delta$-resonance, $\sqrt{<r_\Delta^2>}$ = 0.390
fm, is larger than that of the nucleon. This result is easy
anticipate since the $\Delta$-resonance wave function does not
contain $S_{ij}=T_{ij}=L_{ij}=0$ components, where the potential
(\ref{voct}) is strongly attractive at short range, and thus 
the size of the $\Delta$-resonance is
determined mainly by the weak attraction in the $S_{ij}=T_{ij}=1, L_{ij}=0$
quark pairs as well as by the confining interaction. The bigger size
of $\Delta$ has a well known experimental consequence: the $\Delta
\rightarrow N$  electromagnetic form factor falls off 
 faster than the nucleon elastic one.

When the meson-exchange interaction is switched off, the nucleon
matter radius becomes larger, $\sqrt{<r_N^2>}$ = 0.442 fm.
 This illustrates that there is a 
 soft gap between the scale where chiral physics starts to work and
the scale where confinement is important.

The crucial role of  three body effects can also be seen
from the comparison of the root mean square distance
between quarks in the $S_{12}T_{12} =00$ quark pair in the nucleon, 0.354 fm,
with the same distance in a free scalar diquark, 0.708 fm.
Similarly, the  three body effects
and the antisymmetrization
are responsible
for the fact that the root mean square distance in the
$S_{12}T_{12} =00$ quark pair in the nucleon, 0.354 fm,
is similar to that one in the $S_{12}T_{12} =11$ subsystem in the 
nucleon, 0.387 fm,
while the potential is very different in both cases.
A comparison of the  two numbers above  gives an idea
about how unimportant  clustering is in the nucleon. It can also be
seen from Fig. 4 and Fig. 5 where we show probability
density distributions.

With a pure static ``right triangle'' 3q configuration the relation
between  $<r_{12}^2>$ and $<r_{12,3}^2>$ would be
$<r_{12,3}^2>= \frac{3}{4}<r_{12}^2>$. This relation is
almost exactly satisfied with the $\Delta$ wave function or
with the $3q$ wave function when the meson-exchange 
interaction is switched off.
In the nucleon wave function there is a deviation
from this relation but it  is not large. We thus conclude that 
there is no appreciable clustering in the nucleon.

What is the physical reason for the absense of a significant
clustering? The answer is that there are genuine 3-body
effects and the fermi-nature of quarks do not support
clustering. Indeed, if the quarks, say, with numbers
1 and 2 form a pair  $S_{12}T_{12} =00$, the antisymmetry
of the wave function suggests that there are at the same time
pairs with quantum numbers $S_{13}T_{13} =00$ or $S_{23}T_{23} =00$
(along with other quantum numbers). Thus a strong
attraction acts simultaneously  in all quark pairs
which makes the nucleon compact but not clustered much.
Only a much stronger and ``sharper'' interaction in the
$ST =00$ diquark would lead to an appreciable clustering,
but at a cost that $\Delta - N$ splitting will become
enormous. 

\section{Summary}

Here we summarize our main conclusions.

1. The nonperturbative gluonic interaction between quarks,
e.g. instanton-induced one, which is responsible for the
dynamical breaking of chiral symmetry in  QCD  and thus
explains the $\pi - \rho$ mass splitting, iterated in the $qq$
t-channel implies a meson-exchange picture between constituent
quarks, and through the latter also explains  baryons and nuclear
force. This is a simple consequence of crossing symmetry:
if one obtains pion as a solution of the Bethe-Salpeter
equation in the quark-antiquark s-channel, then one inevitably
obtains a pion-exchange in the quark-quark systems as a result
of iterations in the $qq$ t-channel.

2. Due to (anti)screening effects the implications of this nonperturbative
gluonic interaction in $qq$ systems are drastically different when it
is  iterated only in the s-channel as compared to a more general case,
when it is first iterated in the t-channel, leading to a meson exchange, 
and only after that iterated in the s-channel.

3. The effective meson-exchange interaction in $qq$ systems
does not bind diquarks without an additional confining force and
does not induce any appreciable clustering in the nucleon.

\section{Acknowledgement}

L.Ya.G. thanks D. Diakonov and D.O. Riska for  comments.
He is indebted to KEK-Tanashi and Tokyo Institute of
Technology Nuclear Theory Groups for a warm hospitality.
His work  is supported by a foreign visiting guestprofessorship
program of the Ministry of Education, Science, Sports and
Culture of Japan. Work of K.V. is supported by the
OTKA grant No. T17298 (Hungary) and
by the US Department of Energy,
Nuclear Physics Division, under contract No. W-31-109-ENG-39.

\begin{table}
\caption{Relative distances in $N$ and $\Delta$}
\begin{tabular}{ccccc}
system     & ($S_{12},T_{12}$) & probability 
& $\langle r_{12}^2 \rangle$, fm$^2$ & $\langle r_{12,3}^2 \rangle$, fm$^2$ \\
\hline
Nucleon    & (0,0)  & 0.498 &  0.125 & 0.114  \\
Nucleon    & (1,1)  & 0.498 &  0.150 & 0.094  \\
Delta      & (1,1)  & 1.000 &  0.454 & 0.342  \\ 
Conf. only & (0,0) or (1,1)       &       &  0.585 & 0.441  \\
\end{tabular}
\end{table}

\begin{figure}[tb]
\psfig{file=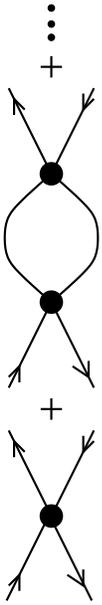,height=8cm}
\caption{The s-channel  ladder in $\bar q q$ system.}
\end{figure}

\begin{figure}[tb]
\psfig{file=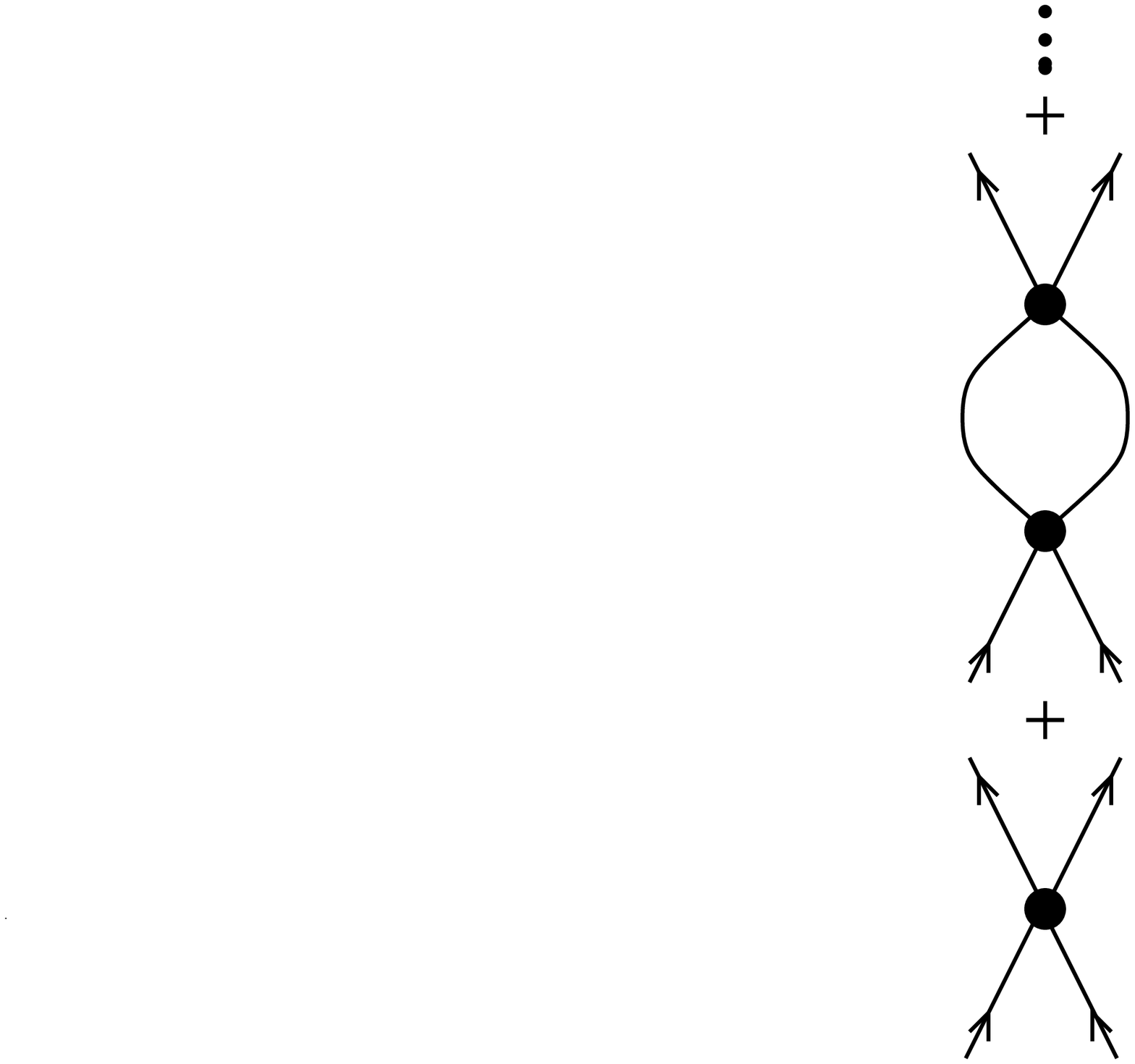,height=8cm}
\caption{The s-channel  ladder in $q q$ system.}
\end{figure}

\begin{figure}[tb]
\psfig{file=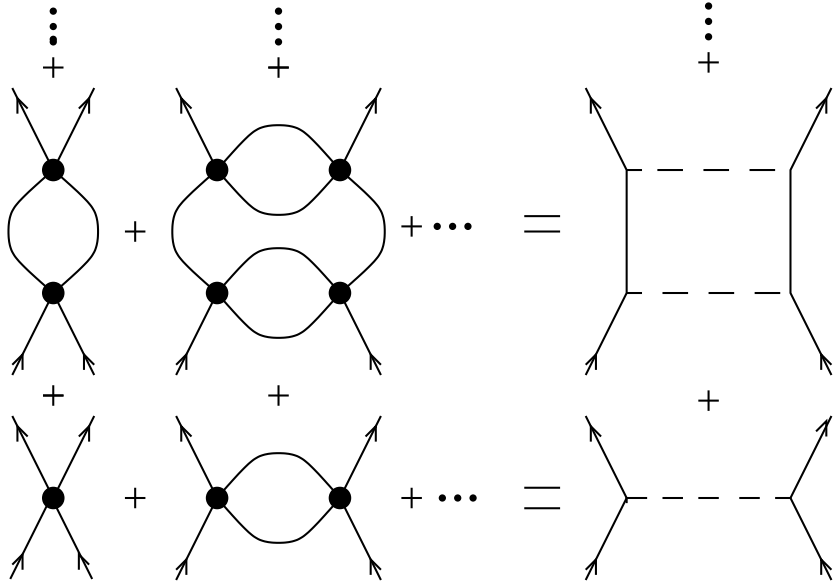,height=8cm}
\caption{The s- and t-channel  ladders in $ q q$ system.}
\end{figure}

\begin{figure}[tb]
\psfig{file=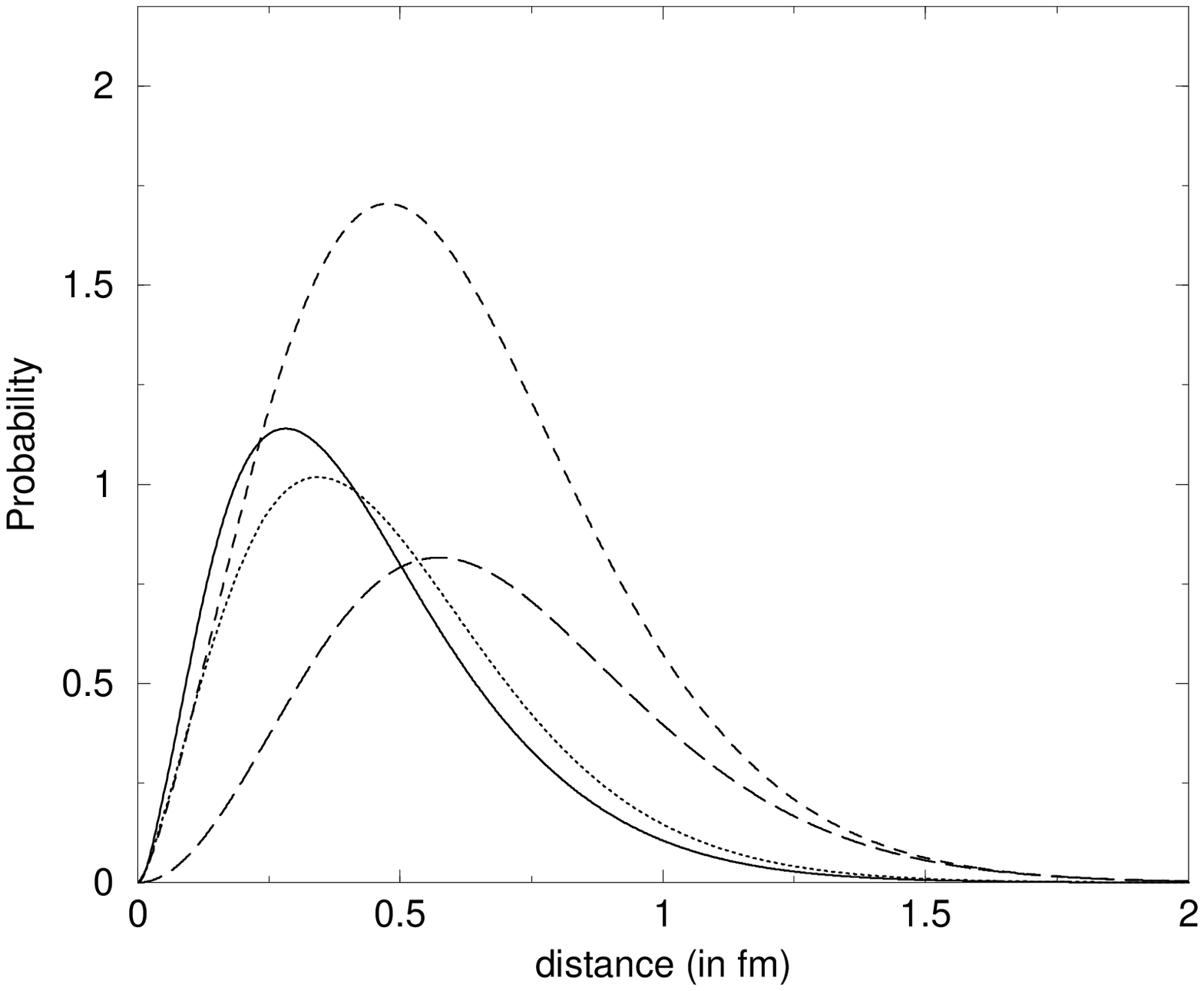,height=8cm}
\caption{The probability distributions (for a definition see the text).
Solid line - $g_{00}(r_{12})$ for $N$;
dotted line - $g_{11}(r_{12})$ for $N$;
dashed line - $g_{11}(r_{12})$ for $\Delta$;
long dashed line - $g_{00}(r_{12}) = g_{11}(r_{12})$
 for $N$  when the spin-dependent
interaction is switched off.}
\end{figure}

\begin{figure}[tb]
\psfig{file=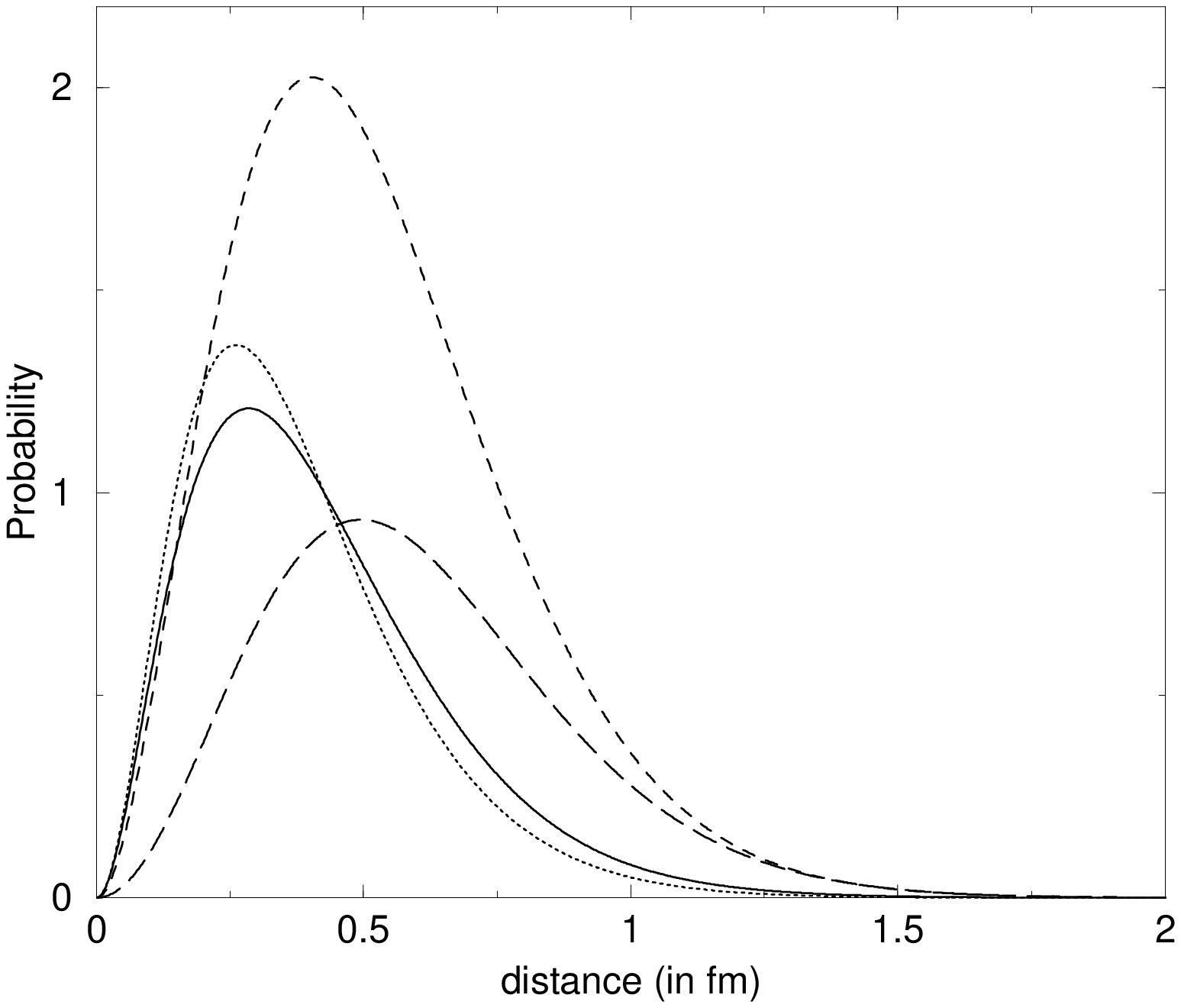,height=8cm}
\caption{The probability distributions (for a definition see the text).
Solid line - $f_{00}(r_{12,3})$ for $N$;
dotted line - $f_{11}(r_{12,3})$ for $N$;
dashed line - $f_{11}(r_{12,3})$ for $\Delta$;
long dashed line -   $f_{00}(r_{12,3}) = f_{11}(r_{12,3})$
for $N$  when the spin-dependent
interaction  is switched off.}
\end{figure}


\begin{references}
\bibitem{RAPP} R. Rapp, T. Sch\"afer, E. Shuryak, M. Velkovsky,
Phys. Rev. Lett. {\bf 81}, 53 (1998).
\bibitem{ALFORD} M. Alford, K. Rajagopal, F. Wilczek,
Phys. Lett. {\bf B422}, 247 (1998).
\bibitem{SHURYAK} 
T. Sch\"afer and E. Shuryak, Rev. Mod. Phys.,
{\bf 70}, 323 (1998), see also references therein.
\bibitem{GR} L. Ya. Glozman and D.O. Riska, Physics Reports {\bf 268}, 263
(1996).
\bibitem{GLOZ} L. Ya. Glozman, Surveys in High Energy Physics, 
{\bf 14}, 109 (1999);
hep-ph/9805345.
\bibitem{GPVW} L. Ya. Glozman, W. Plessas, K. Varga, R. F. Wagenbrunn,
Phys. Rev. {\bf D58}, 094030 (1998).
\bibitem{HES} M. Hes, F. Karsch, E. Laermann, I. Wetzorke,
Phys. Rev. {\bf D58}, 111502 (1998).
\bibitem{VW} U. Vogl and W. Weise, Progr. Part. Nucl. Phys.
{\bf 27}, 195 (1991).
\bibitem{AR} R. Alkofer and H. Reinhard, Chiral Quark Dynamics,
Chapter 5, Lecture Notes in Physics, m33, Springer, 1995. 
\bibitem{BENTZ} N. Ishii, W. Bentz and K. Yazaki, Nucl. Phys.
{\bf A587}, 617 (1995).
\bibitem{DIAKONOV} D. Diakonov, in: Selected Topics in Nonperturbative
QCD, Proc. Enrico Fermi School, Course CXXX, A. DiGiacomo
and D. Diakonov eds., Bologna (1996); hep-ph/9602375, see
also references therein.
\bibitem{NJL} Y. Nambu and G. Jona-Lasinio, Phys. Rev. {\bf 122},
345 (1961); ibid. {\bf 124} 246
\bibitem{THOOFT} 't Hooft, Phys. Rev. {\bf D14}, 3432 (1976);
Erratum: {\it ibid.} {\bf D18}, 2199 (1978).
\bibitem{DFL} D. Diakonov, H. Forkel, M. Lutz, Phys. Lett.
{\bf B373}, 147 (1996).
\bibitem{DPP} D. Diakonov, V. Petrov, P. Pobylitza, Nucl. Phys.
{\bf B306}, 809 (1988).
\bibitem{PREP} L. Ya. Glozman, in preparation
\bibitem{SR}  E.V. Shuryak and J. Rosner, Phys. Lett.
{\bf B218}, 72 (1989).
\bibitem{METSCH} V. H. Blask et al, Z. Phys. {\bf A337}, 327
(1990).
\bibitem{GL} L. Ya. Glozman, hep-ph/9908207
\bibitem{LIU} K.F. Liu et al, Phys. Rev. {\bf D59} 112001 (1999).
\bibitem{NEGELE}  M.-C. Chu, J. Grandy, S. Huang, J. Negele,
Phys. Rev. {\bf D49}, 6039 (1994).

\bibitem{HELM} C. Helminen, Phys. Rev. {\bf C59} 2829 (1999).
\bibitem{BAR} L. Ya. Glozman, in Proceedings of Baryons'98
(Bonn, Sept. 22-26, 1998).

\end{references}
\end{document}